\documentclass[final]{aipproc}

\layoutstyle{8x11single}
\begin{document}

\title 
[Analysis of the Swift Gamma-Ray Bursts duration  ]
{Analysis of the Swift Gamma-Ray Bursts duration  }


\author{I. Horv\'ath}{address={ Dept. of Physics, Bolyai Military University, H-1581 Budapest, POB 15, Hungary}}
\author{L. G. Bal\'azs}{address={ Konkoly Observatory, H-1525 Budapest, POB 67, Hungary}}
\author{Z. Bagoly}{address={ Dept. of Physics of Complex Systems, E\" otv\" os University, H-1117 Budapest, P\'azm\'any P. s. 1/A, Hungary}}
\author{J. Kelemen}{address={ Konkoly Observatory, H-1525 Budapest, POB 67, Hungary}}

\author{P. Veres}{address={ Dept. of Physics, Bolyai Military University, H-1581 Budapest, POB 15, Hungary}}
\iftrue

\author{G. Tusn\'ady}{address={ R\'enyi Institute of Mathematics, Hungarian Academy of Sciences, H-1364 Budapest, POB 127, Hungary}}

\fi

% \copyrightholder{Acoustical Scociety of America}
\copyrightyear  {2008}

\begin{abstract}

Two classes of gamma-ray bursts have been identified in the BATSE catalogs characterized by durations shorter and longer than about 2 seconds. There are, however, some indications for the existence of a third type of burst. Swift satellite detectors have different spectral sensitivity than pre-Swift ones for gamma-ray bursts. Therefore it is worth to reanalyze the durations and their distribution and also the classification of GRBs. Using The First BAT Catalog the maximum likelihood estimation was used to analyzed the duration distribution of GRBs. The three log-normal fit is significantly (99.54\% probability) better than the two for the duration distribution. Monte-Carlo simulations also confirm this probability (99.2\%). 

\end{abstract}

\date{\today}

\maketitle

\section{Introduction}

It has been a great challenge to classify  gamma-ray bursts (GRBs).
 Using The First BATSE
Catalog, \cite{kou93} found a bimodality in the
distribution of the logarithms of the durations. In that paper
they used the parameter $T_{90}$ (the time in which 90\% of the
fluence is accumulated \citep{kou93}) to characterize the
duration of GRBs. Today it is
widely accepted that the physics of these two groups 
 are different, and
these two kinds of GRBs are different phenomena
\citep{nor01,bal03}.
In the Swift database
the measured redshift distribution
for the two groups are also different, for short burst the
median is 0.4 \citep{os08} and for the long ones it is 2.4
\citep{bag06}.

In a previous paper using the Third BATSE Catalog 
\cite{hor98} have shown that the  duration ($T_{90}$) distribution
of GRBs observed by BATSE could be well fitted by a sum
of three log-normal distributions. We find it statistically
unlikely (with a probability $\sim 10^{-4}$) that there are only
two groups.  Simultaneously, \cite{muk98} report the
finding (in a multidimensional parameter space) of a very
similar group structure of GRBs.  Somewhat later several authors
\citep{hak00,hak03,bor04,hak04,chat07} included more
physical parameters into the analysis of the bursts (e.g.
peak-fluxes, fluences, hardness ratios, etc.). A cluster
analysis in this multidimensional parameter space suggests the
existence of the third ("intermediate") group as well
\citep{muk98,hak00,bala01,rm02,chat07}. The physical existence
of the third group is, however, still not convincingly proven.
However, the celestial distribution of the third group is
anisotropic \citep{mesz00,li01,mgc03}.  All these results  mean
that the existence of the third intermediate group in the BATSE
sample is acceptable, but its physical meaning, importance and
origin is less clear than those of the other groups. Hence, it
is worth to study new samples like the Swift data.

In Sect. 2 we discuss the  method used in the paper. In Sect.  3
uni-, bi-, tri- and tetra-modal log-normal fits made by using
the maximum likelihood method are discussed. In Sect. 4 one
thousand Monte-Carlo simulations are shown concerning the
significance of the fits. In Sect. 5 we discuss some further
details. The conclusions are given in Sect. 6.

\section{The method}

There are several methods to test significance. For example the
$\chi ^2$ method which we used in our first paper \citep{hor98}
to analyze the $T_{90}$ distribution of the BATSE bursts is not
useful here, because of the small population of short bursts
in the Swift sample.

In the Swift BAT Catalog \citep{sak08} there are 237 GRBs, of
which 222 have duration information. Fig. 1.  shows the
$\log T_{90}$ distribution. To use the $\chi^2$ method one has
to bin the data. If the number of counts within some bins is small
the method is not applicable. The Maximum Likelihood (ML) method is
not sensitive to this problem, therefore for the (Swift) BAT
bursts the maximum likelihood method is much more appropriate.

The ML method assumes that the probability density function of an
$x$ observable variable is given in the form of $g(x,p_1,...,p_k)$
where $p_1,...,p_k$ are parameters of unknown value. Having  $ N $
observations of $x$ one can define the likelihood function in the
following form:

\begin{equation}
l = \prod_{i=1}^{N}  g(x_i,p_1,...,p_k) ,\
\end{equation}

\noindent or in logarithmic form (the logarithmic form is more
convenient for calculations):

\begin{equation}
L= \log l = \sum_{i=1}^{N} \log  \left( g (x_i,p_1,...,p_k) \right)
\end{equation}

The ML procedure maximizes $L$ according to
$p_1,...,p_k$.  Since the logarithmic function is monotonic the
logarithm reaches the maximum where $l$ does as well.
The confidence region of the estimated parameters is given by the
following formula, where $L_{max}$ is the maximum value of the
likelihood function and $L_0$ is the
likelihood function  at the true value of the
parameters \citep{KS73}:

\begin{equation}
2 ( L_{max} -  L_0) \approx \chi^2_k, \label{eq:chi}
\end{equation}

\section{Log-normal fits of the duration distribution}

Similarly to \cite{hor02} we fit the $\log T_{90}$  distribution using
ML with a superposition of $k$ log-normal components, each
of them having 2 unknown parameters to be fitted with $N=222$
measured points in our case. Our goal is to find the minimum value
of $k$ suitable to fit the observed distribution. Assuming a
weighted  superposition of $k$ log-normal distributions one has to
maximize the following likelihood function:

\begin{equation}
L_k = \sum_{i=1}^{N} \log  \left(\sum_{l=1}^k   w_lf_l(x_i,\log
T_l,\sigma_l ) \right)
%\label{eq:ml}
\end{equation}

\noindent where $w_l$ is a weight, $f_l$ a log-normal function with
$\log T_l$ mean and $\sigma_l $ standard deviation having the form
of

\begin{figure}
\centering
%\resizebox{\hsize}{!}
{\includegraphics[angle=0,width=14cm]{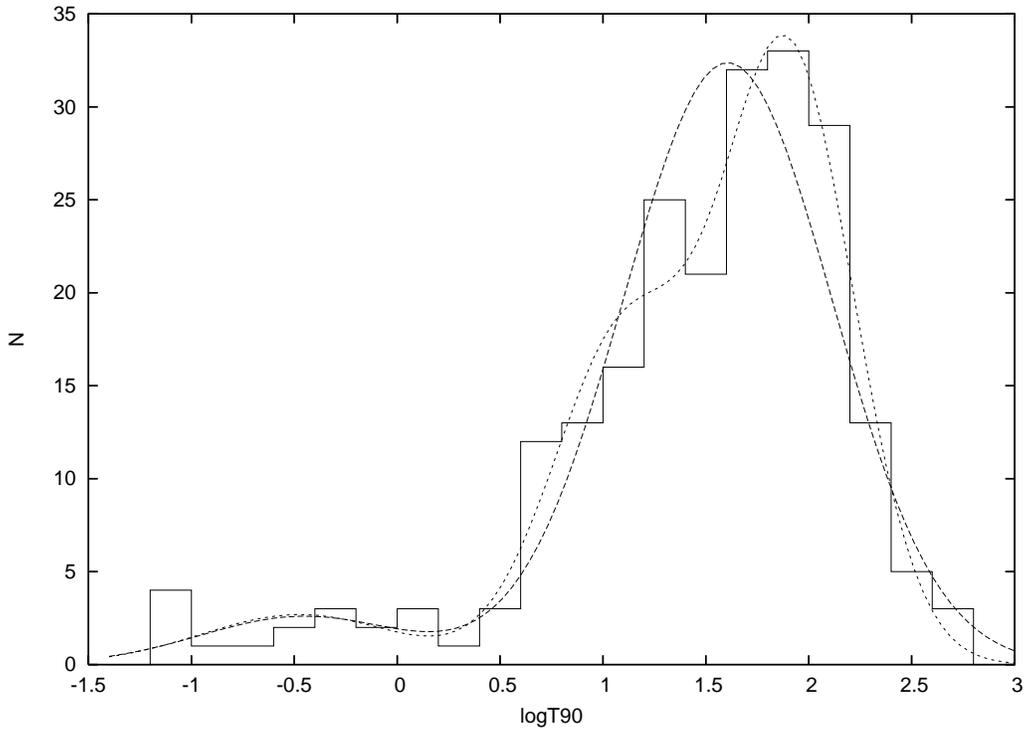}}
\caption{Fit with two and three log-normal component for the duration distribution of BAT bursts. }
\end{figure}

\begin{equation}
f_l = \frac{1}{ \sigma_l  \sqrt{2 \pi  }}
\exp\left( - \frac{(x-\log T_l)^2}{2\sigma_l^2} \right)  
\label{fk}
\end{equation}

\noindent and due to a normalization condition
   $ \sum w_l= N  $.
  We used a simple
C++ code to find the maximum of $L_k$. Assuming only one log-normal
component the fit gives $L_{1max}=951.666$ but in the case of
$k$=2 one gets $L_{2max}=983.317$ with the solution displayed in Fig.~1.

Based on Eq. (\ref{eq:chi}) we can infer whether the addition of a
further log-normal component is necessary to signifincantly improve
the fit. We take the null hypothesis that we have already reached
the the true value of $k$. Adding a new component, i.e. moving
from $k$ to $k+1$, the ML solution of $L_{kmax}$ change to
$L_{(k+1)max}$, but $L_0$ remained the same. In the meantime we
increased the number of parameters by 3 ($w_{k+1}$, $logT_{k+1}$
and $\sigma_{(k+1)})$. Applying Eq. (\ref{eq:chi}) to both
$L_{kmax}$ and $L_{(k+1)max}$ we get after subtraction

\begin{equation}\label{kk1}
2 ( L_{(k+1)max} -  L_{kmax}) \approx \chi^2_3 \, .
\end{equation}

\noindent For $k=1$ $L_{2max}$ is greater than $L_{1max}$
by more than 30, which gives for $\chi^2_3$ an
extremely low probability of $5.88\times 10^{-13}$. This means that the
two log-normal fit is really a better approximation for
the duration distribution of GRBs than one log-normal.

Thirdly, a
three-log-normal fit was made combining three $f_k$ functions with
eight parameters (three means, three standard deviations and two
weights).   The highest value of the
logarithm of the likelihood ($L_{3max}$) is 989.822.  For two
log-normal functions the maximum  was $L_{2max}=983.317$.  The maximum thus
improved by 6.505. Twice this is 13.01 which gives us the
probability of 0.461\% for the difference between $L_{2max}$
and $L_{3max}$ is being only by chance. Therefore there is only a
small chance the third log-normal is not needed. Thus, the
three-log-normal fit (see Figure 1.) is better and there is a 0.0046
probability that it was caused only by statistical fluctuation.

One should also calculate the likelihood for
four log-normal functions. The best logarithm of the ML is 990.323.
It is bigger by 0.501 than it was with three log-normal functions.
This gives us a low significance (80.1\%),
therefore the fourth component is not needed.

\section{1000 Monte-Carlo simulations using the two-component fit}

We can check the 0.0046 probability, which we get for the
maximum likelihood calculation, using a Monte-Carlo (MC)
simulation and adopting the following procedure.  Take the
two-log-normal distribution with the best fitted parameters of the
observed data, and generate 222 numbers for $T_{90}$ whose
distribution follows the two-log-normal distribution.  Then find
the best likelihood with five free parameters (two means, two
dispersions and two weights; but the sum of the last two must be
222). Next we perform a fit with the three-log-normal distribution
(eight free parameters, three means, three dispersions and two  independent weights).
Finally, we take the difference between the two logarithms of the
maximum likelihoods that gave one number in our MC simulation.

We have carried out this procedure for 1000 simulations each with 222
simulated $\log T_{90}$s. There were 8 cases when the log-likelihood difference
was more than the one obtained for the BAT data (6.505). 
Therefore the MC simulations confirm the result obtained by applying Eq. (\ref{kk1}) and give
a similar (0.8\%) probability that a third group is merely a statistical fluctuation.

\section{ Discussion}

It is possible that the fit using three log-normal functions is
accidental, and that there are only two types of GRBs. However,
the probability that the third component is only a statistical
fluctuation is 0.5-0.8 \%.

One can compare the burst group weights with previous results. BAT
sensitivity is different to BATSE sensitivity \citep{fis94,band03}.
BAT is more sensitive at low energies which means it can
observe more X-ray flashes and soft bursts and probably fails to detect many
hard bursts (typically short ones). Therefore one expects more
long and intermediate bursts and fewer short GRBs. In the BAT data
set there are only a few short bursts. Our analysis could only find
16 short bursts (7\%). The robustness of the ML method is demonstrated here
because a group with only 7\% weight is identified.
Previously in the BATSE database intermediate bursts were
identified by many research groups. However, in this class
different frequencies were found  representing 15-25 \% of BATSE GRBs
\citep{muk98,hak00,bala01,rm02,hor06}.

\section{Conclusion}

\begin{enumerate}

\item Assuming that the $T_{90}$ distribution of the short and
long  GRBs is log-normal,  the probability that the third group is
a chance occurance is about 0.5-0.8 \%.

\item Although the statistics indicate that a third component
is present, the physical existence of the third group is still
debatable. The sky distribution of the third component
is anisotropic as proven by \cite{mesz00} and \cite{li01}.
Alternatively \cite{hak00} believe the third statistically proven
subgroup is only a deviation caused by complicated instrumental
effects, which can reduce the duration of some faint long bursts. This
paper does not deal with this particular effect, however the
previously studied BATSE sample shows a similar group structure.
This agreement suggests that the third component is possibly real, not an instrumental
effect (the BATSE detectors and the Swift BAT are different kinds
of instruments).

\item The observed  frequencies in the three classes are different
for BATSE and BAT.  Both samples are dominated,
however, by the long bursts. The short bursts are less populated in BAT than in BATSE but the
intermediate group is more numerous. This is understandable, since BAT is less sensitive
in high energy than BATSE was and more sensitive in low energy and
short bursts are the hardest group and intermediate ones are
the softest. Therefore BAT can observe more intermediate bursts
and much fewer short ones than BATSE did.

\item The existence and physical properties of the intermediate
group need further discussion to elucidate the
reality and properties of this class of GRBs.

\end{enumerate}

\begin{theacknowledgments}
This research was supported in part through  OTKA T048870  grant and Bolyai Scholarship (I.H.).
\end{theacknowledgments}

\bibliographystyle{aipproc}
\bibliography{horv}

\end{document}